# Sensing Shallow Seafloor and Sediment Properties, Recent History

Michael M. Harris, William E. Avera, Andrei Abelev, Frank W. Bentrem and
L. Dale Bibee (Naval Research Laboratory, Stennis Space Center, MS)

Abstract       Near surface seafloor properties are needed for recreational, commercial, and military applications. Construction projects on the ocean seafloors often require extensive knowledge about strength, deformability, hydraulic, thermal, acoustic, and seismic characteristics for locating stable environments and ensuring proper functioning of structures, pipelines, and other installations on the surface of and buried into the marine sediments. The military is also interested in a variety of seafloor properties as they impact sound propagation, mine impact burial, trafficability, bearing capacity, time-dependent settlement, and stability of objects on the seafloor. Point measurements of sediment properties are done using core samplers and sediment grab devices (with subsequent lab analysis) and in-situ probes. These techniques are expensive in terms of ship time and provide limited area coverage. Sub-bottom acoustic and electromagnetic sensors can provide profiles of near surface sediment information with improved coverage rates. Fusion techniques are being developed to provide areal extent of sediment information from multiple sensors. This paper examines the recent history of techniques used to measure sediment properties in the upper portions of the seafloor and in shallow (<100m) water.

## I. BACKGROUND

The IEEE Ocean Engineering Society (OES) has seventeen technical committees covering everything from unmanned submersible vehicles to space remote sensing. The mission of the Oceanographic Instrumentation Committee is to provide a technical forum to make the results of research and development in oceanographic instrumentation known to the public.

The OI committee focuses on sensors and systems that measure the following water and sediment properties: conductivity; temperature; depth; pressure; salinity; sound speed; water samplers; wake measurements; chemical properties; seafloor properties; sediment properties; non-acoustic communication; and navigation and positioning. Interests include new sensor developments; sensor performance comparisons; and applications characterizing the ocean environment. In concert with the 400$^{th}$ anniversary of Quebec and the 40$^{th}$ anniversary of the IEEE Oceanic Engineering Society, this paper examines recent history of sensor technologies and measurement techniques used to quantify seafloor and sediment properties.

## II. APPLICATIONS

Civilian and military communities need to determine seafloor and sub-bottom structure to predict geotechnical and geoacoustic properties in the upper few meters of the seafloor. This information is used for seafloor-engineering applications, trafficability estimates, and as input to acoustic propagation models[1].

On the civil side, seafloor engineering is required for the development of offshore resources and supporting coastal infrastructure. Information on surficial marine sediments is used for: cable and pipelines route surveys; anchors and moorings; pilings and footings; placing items on the seafloor; dredging; and break water, jetties, and harbors construction. Design and placement of pilings, platform footings, anchorages and moorings depends on sediment properties as does the mobility of equipment designed to traverse the seafloor. Knowledge of the seafloor is also required for salvage operations and to perform marine archeology studies.

On the Military side trafficability, impact and subsequent burial, and acoustic propagation in the upper sediments are major considerations in addition to seafloor construction. Seafloor and sediment characteristics affect reflection and absorption of acoustic energy from sonars used to look for targets in the water column, at the sediment water interface, and below the seafloor[2]. Similarly, sediment characteristics affect acoustic positioning sonars used to track vehicles. Sediment properties also determine impact burial for objects and mines, and subsequent burial due to wave and current action.

Sediment properties can be segmented into categories that describe individual grains, compositional properties and bulk properties. A combination of direct and remote sensing technologies can be used to measure desired properties. Table 1 cross references sediment properties with sensing techniques. The authors understand that there are more properties and more techniques that can be listed; however, the table is presented as a summary in the spirit of the history of seafloor sensing techniques. In the table, if the intersection block of a property and technique is grey the property is directly measured. If the intersection is dotted, the property is empirically determined. Historically speaking, the newer sensing techniques appear to the right.

| Sediment Properties | Sampling | | Measurement Techniques | | | | | | | | | | | | | | |
|---|---|---|---|---|---|---|---|---|---|---|---|---|---|---|---|---|---|
| | | | Direct Measurement (Contact) | | | | | | | Remote Measurement (Non-contact) | | | | | | | |
| | | | Dynamic Probes | | Static Probes | | | | | | Acoustic | | | | EM | | |
| | Cores | Grabs | XBP | STING/ESP | Vane | Cone Penetro-meter | Acoustic | Thermal | PMT, DMT | Piezo-meter | Profiling | Multibeam | SSS | Chirp | Resistivity | CSEM | AEM |
| **Compositional Properties** | | | | | | | | | | | | | | | | | |
| mineral composition | ▓ | ▓ | | | | | | | | | | | | | | | |
| organic matter content | ▓ | ▓ | | | | | | | | | | | | | | | |
| grain shape | S | S | | | | | | | | | | | | | | | |
| **Bulk Properties** | | | | | | | | | | | | | | | | | |
| *GeoTech* grain size distribution | ▓ | ▓ | | | | | | | | | | | | | | | |
| mean grain size | ▓ | ▓ | | | | | | | | | | C | | | | | |
| liquid and plastic limits | C | C | | | | | | | | | | | | | | | |
| density | ▓ | ▓ | | | | | | | | | ▒ | ▒ | | ▒ | | | |
| relative (min, max) | S | S | | | | | | | | | | | | | | | |
| specific gravity | ▓ | ▓ | | | | | | | | | | | | | | | |
| moisture content | ▓ | ▓ | | | | | | | | | | | | | | | |
| porosity | ▓ | ▓ | | | | | | | | | ▒ | ▒ | | ▒ | ▒ | ▒ | ▒ |
| shear strength | ▓ | ▓ | S | C | C | C | | | ▓ | | | | | | | | |
| deformability | | | S | | | | | ▒ | ▓ | | | | | | | | |
| permeability | | | | | | | | | ▓ | ▓ | | | | | | | |
| stress state - in-situ | ▒ | | | | | | | | ▓ | | | | | | | | |
| pore pressure - in-situ | | | | | | | | | | ▓ | | | | | | | |
| *Acoustic* surface texture (>.3m) | | | | | | | | | | | | | ▓ | | | | |
| impedance | ▓ | | | | | | ▓ | | | | ▓ | ▓ | | ▓ | | | |
| comp. & shear velocities | ▓ | | | | | | ▓ | | | | ▒ | ▒ | | ▒ | | | |
| acoustic attenuation | ▓ | | | | | | ▓ | | | | ▒ | ▒ | | ▒ | | | |
| acoustic roughness | ▓ | | | | | | | | | | | | ▓ | | | | |
| *EM* electrical conductivity | ▓ | | | | | | | | | | | | | | ▓ | ▓ | ▓ |
| *Thermal* thermal conductivity | ▓ | | | | | | | ▓ | | | | | | | | | |

KEY
▓ Direct Measurement   S Granular materials, i.e. silts, sands, gravels
▒ Derived/Empirical Property   C Cohesive materials, i.e. clays, muds

XBP - eXpendable Bottom Profiler
STING - Sea Terminal Impact Naval Gage
ESP - Electronic Sediment strength Probe
SSS - Side Scan Sonar
CSEM - Controlled Source EM
AEM - Airborne Electromagnetics
DMT - Dilatometer
PMT - Pressuremeter

**Table 1**, Sediment Properties and Measurement Techniques

The civil, recreational or military application will determine what properties should to be measured. The more commonly measured bulk properties are shear strength, impedance, shear velocity and acoustic attenuation. The dominance of these measurements makes sense based on general geotechnical and geoacoustic requirements. Geotechnical parameters such as shear strength determine how a foundation or anchor will perform. Impedance is used as an input to acoustic propagation models.

## III. SAMPLING TECHNIQUES

*Classification*

Sampling methods in the marine environment can generally be grouped according to several principles. They can be intended for shallow or deep sampling into the sediment bed – from small and shallow grabs to cores as deep as 20m; by the degree of material structure disturbance during the sampling procedure – from totally remolded grab samples to lined corers with relatively intact internal structure; by which material they are mostly applicable to (*e.g.* granular, cohesive), and by the maximum water depths they can be used in. Samplers also differ by the size and geometry of the retrievable specimen – from small hand-driven cylindrical cores able to sample up to 40-50 cm in the softest soils and small grabs probing the top 10-20cm of sediment to large box corers, usually square in cross-section, and long (up to 20m) cylindrical piston corers. Subsequently, these different samplers have vastly varying requirements for the applicable platform to be deployed from, some requiring high-capacity lift winches and high vertical lift booms to manipulate them on deck and deploy.

Most geotechnical investigations require undisturbed specimens, mandating therefore the use of corers of one type or another. Sampling depth requirements can vary widely and are dependent on the specific problem at hand. Majority of investigations will utilize gravity corers, vibratory corers, and piston corers to attain the sufficient sampling depth and recover specimens for laboratory testing of relatively intact structure. Table 2 lists the type of equipment to use for the desired depth of sample. The following is a brief list of several common samplers and corers:

1. Corers: Hand corers, Alpine gravity corer, Benthos gravity corer, Boomerang, Box corers, Piston corers, Phleger, Kajak Brinkhurst, and Vibratory
2. Grab samplers: Birge-Ekman, Petersen, Ponar, shipek, Smyth-McIntyre, and van Veen

Most grab samplers, whether large or small, are based on similar operating concepts and consist of one or more rotating containers, triggering a closure by a certain mechanism when in contact with the sediment floor. Fig. 1 shows examples of these commonly used samplers. As is obvious from their operating principles, sediment structure is generally disrupted upon sampling yielding an averaged representation of the surficial sampling depth of 10-30 cm. In some instances, larger samplers, *i.e.* box corers, can be further subsampled with small push-in tubes, yielding relatively undisturbed specimens.

Corers (Fig. 2), on the other hand yield relatively undisturbed specimens. They differ mainly in size (diameter and sampling depth), bottom sediment catcher mechanism, top core flow valve, and the presence and design of a piston. The valves play a significant role in preserving the integrity of the sediment within the sampling tube during retrieval. Generally, common gravity corers can retrieve up to 5m (in softer sediments) of soil, whereas the longer piston corers can yield as much as 20m of sediment.

Vibratory corers are used in primarily cohesionless granular materials where gravity corers can penetrate to only limited distance and/or unable to retain the sediment to a sufficient amount due to low cohesion with the sampling tube that can be generated. Vibro-corers are similar to the gravity corers but have a motorized unit added to the top of the assembly, generating an oscillating motion that assists in tube's burial in the sediment. Several types are mostly distinguished by the actuator type: pneumatic, hydraulic, or electro-pneumatic. Most are limited to about 200m water depths due to difficulty of providing power to the unit. Maximum penetration can be up to10-12m, unless very soft soils are encountered. Some more innovative techniques can be represented by the Boomerang corer. This is a small free falling sampler (total weight: 85 kg, sampling tube: 120cm long and 6.7cm in diameter) and can be deployed from any small boat. A disposable part of the corer consists of a nose cutter, a core barrel and weights.

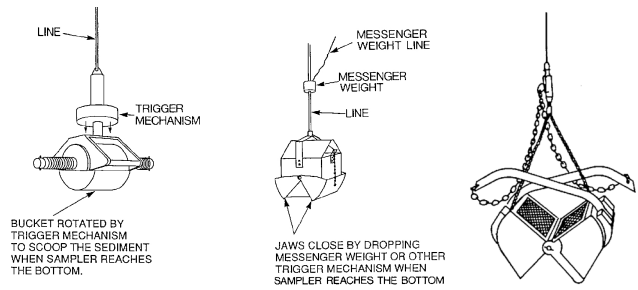

**Fig. 1 Shipek, Birge-Ekman, van Veen samplers**

*Table 2 Sampling depth and equipment[3]*

| Sampling depth | Sampling equipments |
|---|---|
| 0 – 10 cm | Lightweight, small volume grabs (*e.g.* Birge-Ekman, Ponar, Shipek) |
| 0 – 30 cm | Heavy, large-volume grabs (e.g. van Veen, Smyth-McIntyre, Petersen) |
| 0 – 1 m | Single small gravity cores (e.g. Kajak-Brinkhurst, Phleger, Alpine), box-cores, multiple cores |
| 0 – 5 m | Single large gravity corers (e.g. Benthos) |
| > 5m | Piston corers (e.g. Norwegian DWS – Deep Water Sampler) |

The retrievable part consists of the core barrel liner tube and two spherical floater spheres that are released after the penetration into the sediment and float back to the surface pulling the liner tube with the sample with them. Reported sampling water depth can be up to 9000m.

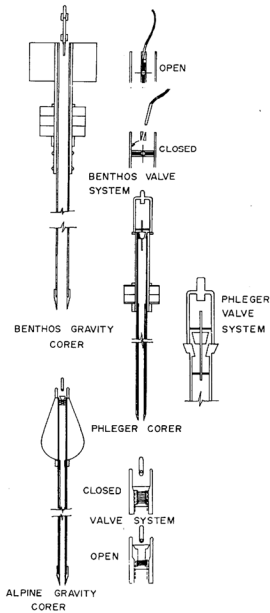

For sampling depth greater than those that can be attained by the piston corers or in case of hard granular materials or even very stiff clays with possible bolder inclusions where surface corers have difficulty penetrating, down-hole sampling can be employed in conjunction with rotary drilling. Here, the drill-hole is advanced to a desired depth, then the drill head is lifted and a sampler is deployed through the drillstring. Then the drilling can continue to the next depth of sampling interest. Drillstring samplers include wireline percussion, latch-in push, and hydraulic piston samplers. While these samplers can be lowered into the unstabilized drillstrings, stabilization can allow for higher quality sampling as well deployment of a variety of in-situ probes.

*Analysis of specimens*

Ranges of tests and material properties that can be determined from different sampling methods are wide and with significant differences among them. Grab samples, representing mostly disturbed material from the very surface of the sediment are used primarily in characterization of material grain-size characteristics, mineral composition, and organic matter content. Other soil structure-independent parameters, such as grain-shape descriptors for coarse materials and plasticity characteristics (liquid and plastic limits) for fine clayey soil can also be determined from grabs, as well as various environmental descriptors.

Most geotechnical laboratory tests that are conducted on retrieved specimens require that the undisturbed structure of the soil is preserved as much as possible. These tests are geared to identifying complex material response and characteristics defining strength, deformability, hydraulic, thermal behaviors, acoustic response, material stress history and anisotropy.

**Fig. 2 Common corers**

*Trends*

Most innovations in sampling relate to minimization of specimen disturbance during penetration and retrieval (especially in very soft sediments). In general, the current techniques are quite mature and are generally considered adequate, thus, not resulting in many significant new developments. Most grab samplers are similar to one another in construction and operating principles. Improvements or changes, if done, have to do mostly with reliability of trigger mechanisms and prevention of material wash-off during retrieval. In surface coring technology, similar attempts to minimize sample disturbance during penetration and retrieval are done, mostly focusing on cutting shoe, bottom retainer, and valve/piston modifications. One example of relatively recent developments could be the Norwegian Deep Water Sampler (DWS). This is a heavy piston coring system that attempts to minimize specimen disturbance by all the modifications listed above as well as instrumenting the piston to monitor the amount of suction developed during lift. The corer is incorporated into a frame that is lowered to the seafloor with the sampling unit then pushed into the sediment. Sampling depths of up to 20m are reported. These trends may be indicative of future possible developments in the marine sampling techniques, including deeper sampling and increased instrumentation of corers, allowing for greater information collection on influences and changes in the specimen conditions during sampling and retrieving. Pressurized sampling may be another avenue of future work, allowing for sample retrieval while maintaining the ambient hydrostatic pressure.

## IV. MEASUREMENT TECHNIQUES AND SENSOR TECHNOLOGY

In general, determination of sediment properties of marine sediments can be achieved in laboratory environment on-board a vessel, in an on-shore lab, or in-situ. Field (or in-situ) testing techniques can be grouped into two major classes, those utilizing direct or contact methodology and those using indirect or remote approaches. Contact methods include direct measurements on the sediment surface (or in the down-hole mode) by static instruments, lowered from a vessel or dynamic probes, deployed in free-fall through the water column.

*Direct Measurement Techniques*
   *In-situ testing: static probes*
   Contact measurements in-situ are typically performed in one of the two configurations: using a remote platform or through a drillstring (Fig. 3). Different organizations developed a number of tools for this purpose: Swordfish, Stingray, and Dolphin (originally McClelland Engineers), Wipsampler , Seaclam, Seasprite, and Seacalf (Fugro), among others. Several probes, *i.e.* cone penetrometer, require the use of a stabilized drillstring, where all vertical motions of the vessel are suppressed and the vessel is providing a controlled thrust for the probe. Remote tethered platforms are generally more economical but do not provide drilling capability and are limited to about 200kN of thrust. Examples of such platforms are Stingray (formerly, McClelland) and Seacalf (Fugro). Examples of several down-hole samplers and probes are given in Fig. 4 for the Fugro's Seaclam (Wison) drillstring system. A number of smaller platforms are also used for testing surficial sediments with limited probe penetration depths that do not require a large vessel for deployment.

Most commonly used probe types for in-situ testing of soils are the following:
- cone penetrometers;
- vane shear devices;
- pressuremeter/dilatometer/hydraulic fracture systems;
- dedicated compressional and shear wave probes;
- heat flow probes.

**Cone penetrometers** are perhaps the most widely used tool for in-situ investigations and have evolved over several decades of use in a number of important ways. CPT was first introduced in 1934 in the Netherlands (Dutch Cone Penetrometer). Electrical sensors were first developed in 1948 but were not used widely until the 1960s. By the 1980s, the CPT has become widely used by the industry with more subsequent research into various sensor packages. Currently, the cone penetrometer series include the standard CPT, CPTu (piezocone penetrometer), SCPTu (seismic piezocone penetrometer), and electric conductivity cone. More recently, gamma-ray sensor has also been added to some probes, yielding a continuous profile of the sediment density, correlated from the gamma-ray attenuation. All of these cone probes can be advanced using the same equipment. Heaviest remote platform-operated CPT probes can be deployed in up to 6000m water depth (Figro's Seaclaf) with maximum reaction force of 200kN and maximum penetration of 50m. Intermediate-size systems (*e.g.* A.P.v.d.Berg's ROSON) can be deployed in up to 500m depths and provide reaction force of 100kN and penetrate up to 15m into the sediment.

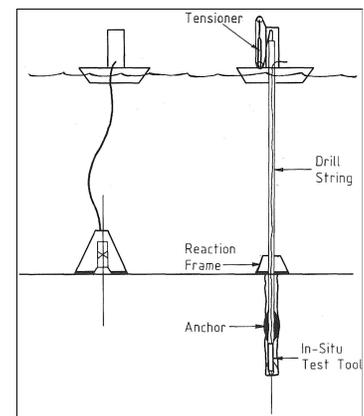

**Fig. 3 In-situ testing systems**

Standard cone penetrometers could be of electric or mechanical type, with the difference being only the mode of operation and the sensor package. Same two quantities are measured for both systems: cone tip resistance and sleeve resistance. These quantities relate to the plastic flow that develops around the tip of the penetrating probe as well as developed friction on the side of the probe, immediately behind the tip. These quantities describe the strength and deformability characteristics of the soil, including friction angle and undrained shear strength, depending on the material.

Piezocone, CPTu (*e.g.* Fig. 5), is equipped with a pressure transducer, in addition to tip and sleeve force transducers. This record yields pore pressure signatures as a result of cone penetration in soils and is used in many correlations with material properties, from assisting in sediment classification to permeability and consolidation, and some stress-history characteristics. Main concern with CPTu deployments is typically assuring full saturation of the porous stone near the pore pressure transducer. This is not always simple to achieve and incomplete saturation can yield highly unreliable results on pore pressure built-up and dissipation. It is possible that more technological and procedures development are needed in this case to simplify operations and avoid errors in measurements.

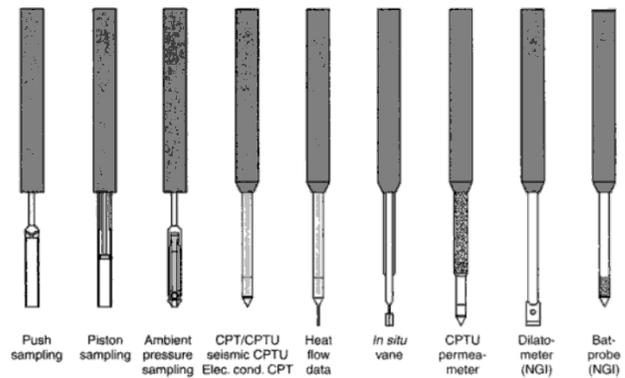

**Fig. 4 Sampling and testing probes for Seaclam (Fugro) system ("Bat" – electrical conductivity probe)**

Additionally, SCPTu is also equipped with either geophones or accelerometers and is able to measure the shear wave velocity generated from a source at the sea bottom and is also able to operate as a regular cone penetrometer. In some cases it may also be able to measure P-waves. Measurements of average shear (or compressional) velocity allows for computation of small strain shear (or bulk) modulus, in addition to the cone, sleeve, and pore pressure transducer-derived parameters.

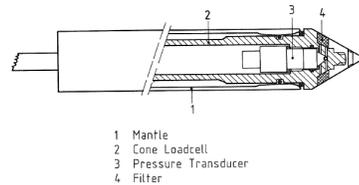

**Fig. 5 Piezocone**

**Vane shear test** (*e.g.* Fig. 6) is performed primarily to obtain the undrained shear strength of soft cohesive deposits. As the CPT it can also be deployed in the drillstring or on a remote platform and provides measurements of peak resistance as well as residual strength, typically after sufficient remolding of the soil has taken place as a result of the vane rotation within the soil.

**Pressuremeter** is a device for measuring lateral deformation and strength properties of soil. It is built for pre-drilled shafts as well as in a self-boring form and is deployed in a drillstring or remote platform configurations. (Hydraulic fracture testing probe is a similar tool intended for shafts drilled in rock formations.) The probe consists of one or three inflatable cylindrical diaphragms, instrumented to monitor and control hydraulic pressure and volume of hydraulic liquid in each one. It is lowered into the drillshaft (or advanced into the soil, if self-boring) and then inflated, recording the pressure in the measuring section *vs.* lateral expansion of the shaft (and the soil). The tool is widely used in terrestrial applications and has seen an increase in marine use over the last 10-20 years. Primarily, it is used for determinations of in-situ horizontal stresses in the soil. Additionally, it can be used to determine undrained shear strength of clays and friction angle of sands as well as shear modulus from unload-reload portions of the pressure-volume curves, and general stress-strain behavior. 'P-Y' (or load-displacement) curves for lateral pile design can also be generated using a pressuremeter.

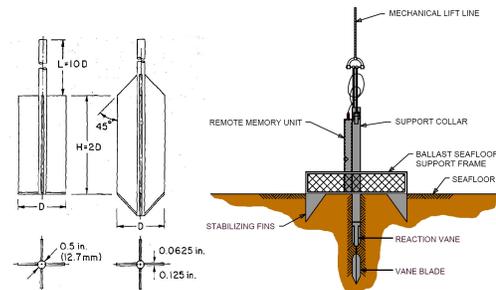

**Fig. 6 Vane shapes and Halibut (Fugro) remote system**

**Dilatometer** is flat spade-like plate (95mm wide, 14mm thick) with an inflatable circular steel diaphragm on one side (60 mm in diameter). Originally designed by Marchetti[4] it has been introduced into the field of offshore testing in the last decade. It is used for mostly same purpose as the pressuremeter and allows for similar derivations of data regarding soil strength, stiffness, and stress-history. Most recent developments introduced a seismic sensor into the probe to allow for the shear wave velocity measurements as well. One may expect similar evolution for this device in the future, if indeed the industry increases its use, including incorporation of additional sensors as occurred with the cone penetrometer technology, yielding additional information about the soil.

**Dedicated acoustic systems** are designed to provide compressional and shear wave characteristics of the sediment. NRL ISSAMS[5] (In-situ Sediment geoAcoustic measurement System) is an example of such a system, designed for measurements in surficial (0-50cm) sediments and water depths of up to 300m. Additionally, data on water temperature, salinity, depth, and a video of operations are also taken. ISSAMS has been used at 83 sites worldwide and is mostly a research tool. In practical geotechnical offshore work, acoustic properties of sediments are more often measured using either remote acoustic sensing or borehole tools incorporated together with other probes, *e.g.* cone penetrometer or dilatometer.

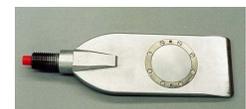

**Fig. 7 Dilatometer**

**Heat flow** probes are designed to study thermal properties of the sediment and consist of a sensor string (typically 1cm diameter) with a series of thermistors and heater elements, attached to the side of a strength element (Fig. 8), generally 3-5m long. The strength element is a steel tube with an added instrumentation package near the top for sensor data acquisition and control storage. Measurements include monitoring heat gradients as a result of the probe insertion as well as heat dissipation with time produced by an impulse of the heater elements. Main parameters of interest include geothermal gradients and thermal conductivities of the surficial 3- to 5-m portion of the seafloor sediment. These data are used for both engineering design of structures under thermal load, such as wellbores and pipelines, as well as sedimentary basin hydrocarbon maturation studies for exploration purposes.

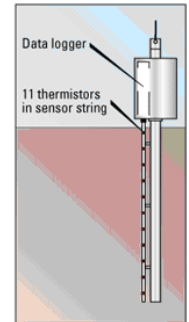

**Trends in static in-situ probe development** included several directions in the recent decade. One trend involves improvements made to the existing probes mostly due to better electronics packages, data acquisition and processing. Another involves incorporation of multiple sensor packages into the same probe technology, as is evident from the history of CPT evolution. Modern probes include different types of measurements in a single unit, *e.g.* tip resistance and sleeve friction cones, pore pressure cones, seismic (p- and s-waves), pressuremeter cones (full-displacement pressuremeter), vibrating cone for liquefaction evaluation, lateral stress cone (for pile analysis and $K_0$ evaluation), logging cone (for nuclear density estimates), resistivity cones (for porosity evaluation), and cones with sampling capabilities (for environmental work). Generally, CPT technology remains one of the most widely used and developed, with decades of experimental and technological improvements as well as related analytical tools and empirical correlations for a wide variety of geotechnical problems. One may observe a beginning of a similar trend with the dilatometer probes, with recent incorporation of the shear wave profiler into the package. Their use, however, appears to be quite limited at this time in marine applications.

**Fig. 8 Heat flow probe**

Additionally, new probe configurations have been introduced over the last decade, *i.e.* full-flow penetrometers for soft or very soft soils. These include ball and T-bar penetrometers. These allow for development of more accurate processing algorithms for extraction of sediment strength parameters, such as undrained shear strength due to the different geometry of the plastic flow developing around the probe tip during insertion and retraction. This geometry allows for more accurate analytical and numerical techniques to be applied, with fewer assumptions, to determine strength characteristics of the sediment.

*In-situ probes: dynamic probes*

Dynamic marine probes are intended for similar purposes as the standard static devices described above, but are typically less sophisticated, perform fewer measurements and are able to probe only to the very limited sediment depth of up to 3m in the softest soils. These can be further categorized into expendable and retrievable probes. Development of all of these, and especially of the expendable probes, are of interest for the military applications, as large dedicated geotechnical platforms are not always appropriate for the military use. All of the probes dicussed below are intended for mostly soft cohesive sediment profiling, where impact penetration and burial can be achieved as a result of a free-fall through the water. In granular sediments, impact burials are rather small and the data produced from such drops are typically unusable.

Main probe types in use today are:
- XBP - eXpendable Bottom Profiler[6];
- STING (Sea Terminal Impact Naval Gage)[7] and ESP (Electronic Sediment Profiler)[8];
- Probos[9];
- Combined systems.

**XBP** was first introduced in the 90s and was designed for rapid surveying of the topmost sediment layer in soft sea bottom sediments. It is a small expendable probe (modified XBT – eXpendable BathyTermograph) that has seen some limited use, mostly in a research framework, by several organizations, including NRL. It's a simple system with a single-axis accelerometer instrumented body connected via an electrical line to the on-board data acquisition and storage system. It generally yields only classification information about the sediment (at most top 30cm), grouping the material into one of four categories according to the maximum deceleration encountered during impact. Some recent numerical modeling advances have allowed for a more direct material parameter extraction (undrained shear strength) from the deceleration time history, with some assumptions about strain-rate dependency. Some additional work is being done at NRL in the same area, with possible better algorithms to follow. This tool, however, suffers from small size and mass, allowing for only very shallow penetration of the surface. Additionally, it is only produced by a single manufacturer and its cost is apparently another hindrance to its wider use and acceptance. Additional developments in the software algorithms, electronics and data acquisition could perhaps increase the applicability and use of this tool for rapid surficial sediment investigations.

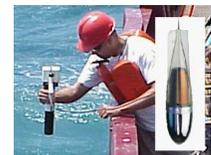

**Fig. 9 XBP probe**

**STING** probe, similarly to the XBP, measures the deceleration profile as a result of a free-fall through the water and sediment penetration. Unlike XBP, this is a recoverable tool with the self-contained data acquisition block located in the main fin-stabilized body (weighing 10kg).

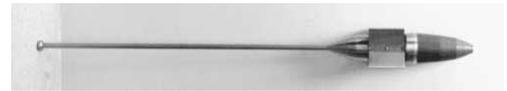
Fig. 10 STING (with 1m rod & 25mm foot)

Steel rods (1-3m long) are attached to the probe together with one of four disks at the end of the rods. These discs range from 25 to 70 mm in diameter. The probe is allowed to free fall and then lifted from the sediment (typically several times in same series), after which, the probe if recovered, connected to a computer, and the drop data downloaded and analyzed. In addition to the single-axis accelerometer, a water pressure transducer is also housed in the probe's body. Recorded water pressure is used for velocity and water depth calculations. Current processing algorithms are rather crude yielding different values of the 'bearing strength' for different end disks used. NRL is currently working to improve these algorithms in accuracy as well as yield parameters of greater engineering meaning *i.e.* undrained shear strength, as opposed to the rather ambiguous 'bearing strength' values. Advantages over XBP, however, include larger weight and longer (up to 3m) penetration depths. Electronic Sediment Probe (ESP) is a very similar device, with a 1m long rod that has an almost identical configuration and design. It has been used by the Australian Navy for sediment characterization.

**Probos** is yet another STING-like instrument (modeled after it) that is currently in research and development stages that improves the design of the model-probe by incorporating a cone-tip load cell. This addition significantly improves the ability to process and analyze the drop records, including utilization of the CPT experience and algorithms where cone force is also measured. As an experimental device, it has seen only very limited use but it holds promise of greater accuracy in deducting parameters such as undrained shear strength.

**Combined systems** may be represented by an FFCPT[10] (Brooke-ocean). This is a heavy (52kg) dynamic probe that includes several sensors: accelerometer (for correlations with undrained shear strength, Su), pore-pressure transducer (near the tip, for Su correlations), electrical resistivity module (for correlations with porosity, which may be reasonable in granular materials but are inaccurate for clays), and an optical backscatter sensor near the tip for surface detection in particularly soft muds (e.g. fluid muds). It is also equipped with a tail pressure transducer for hydrostatic water pressure measurements and decent velocity calculations. The probe can penetrate up to 3m, in soft sediments. The device is still under development and is used in combination with an automatic winch system for continuous deployment and retrieval while underway.

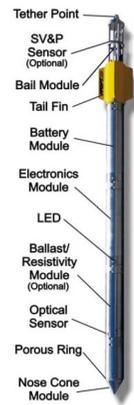
Fig. 11 FFCPT

**Trends in dynamic in-situ probe development** are somewhat hard to identify as these devices see generally very limited use, almost exclusively being the domain of research groups. The demand for these probes at the current stage of the technology development appears to be small, including military applications. It appears that additional sensor packages for retrievable devices are a trend, whereas expendable probes have not seen much development at all in the last 10 years and are mostly limited to minor data acquisition and processing improvements.

*Laboratory testing*

In general there are two types of laboratory testing: on-board a vessel and in an on-shore lab. Most tests can only be performed in an on-shore laboratory with only a few limited investigations done in on-board conditions. These often include specimen classification and simple tests such as miniature vane tests (yielding undrained shear strength) and fall-cone tests (correlating with undrained shear strength and liquid limit). It is possible that more wide-spread use of core-logging on-board vessels can be expected, yielding density (from gamma-ray attenuation) and p-wave velocity.

Most geotechnical investigations conducted today make extensive use of the undisturbed cores and are done in an on-shore laboratory. Detailed review of these testing methods and the types of material parameters that can be inferred from them are beyond the scope of this paper. Below is a brief list of main testing categories, performed routinely in geotechnical laboratories:
- hydraulic conductivity – falling or constant-head tests;
- strength and stress-strain behavior:
  - oedometer tests, yielding consolidation and creep parameters, including coefficients of consolidation, secondary compression, compression/expansion index, pre-consolidation pressure, OCR, and others;
  - direct and simple shear tests, yielding friction and dilation angles;
  - triaxial tests, including, most commonly unconfined compression, unconsolidated undrained, isotropically consolidated undrained/drained, $K_0$-consolidated undrained/drained tests, tests in compression, extension, or cyclic, yielding many strength and deformability parameters;
  - true-triaxial tests, including cubic triaxial (rigid-, flexible-, or mixed-boundary configurations), yielding detailed 3D stress-strain-strength descriptions;
  - torsional cylinder tests, including solid and hollow cylinders, yielding stress-strain-strength as well as stress-rotation analysis of soil behavior;
  - and thermal conductivity tests for coupled thermo-mechanical problem parameterization.

Some of the trends in the last decade has been the much more commonly accepted use of the small-strain stiffness testing as well as multi-axial testing, with hollow-cylinder and cubic triaxial testing systems now available commercially. Additionally, CT (computed tomography) is applied much more widely in research and some industrial settings for the description of natural granular materials, including some marine sediments. It is possible that the use of these advanced laboratory testing methods will continue to garner wider acceptance and become more wide-spread with advancements in technology as well as reduced costs of these systems.

*Remote Measurement Techniques*
   *Magnetic and Electric Techniques*

Electric and Electromagnetic (EM) sensor technology have been used for more than 80 years to explore the surface and shallow sub-surface of the earth. EM sensing technology exploits the electrical or magnetic properties to expose differences in the shallow subsurface materials and these differences are used to identify or focus on specific geologic sediment types. Both electric and EM sensors technologies measure the basic electrical properties of the sediment, but with different sensor types, to different depths, or in different frequency regions. For bottom sediments, this is mainly the electrical conductivity. There are many types and configurations of electric, magnetic, and EM sensor technologies. This section focuses on a few examples. A quick explanation of the sensor technology will help to associate the differences in the technology.

Electric sensors use electrodes to measure the electrical conductivity of a material. This often includes a set of source electrodes to put current into the shallow bottom and receiver electrodes to measure the resulting voltage potentials. In this configuration, geometry is very important because the current flows between the source electrodes and spreads out through any conductive path. Measures of the resulting electric potentials are taken with the receiver electrodes and the measure depends on the location relative to the source electrodes. Since the sea bottom is not electrically uniform, a simple model is often used to estimate the bottom properties. In the simplest case this would be a uniform halfspace model.

EM sensor technology measures similar properties to the electric method using a different approach. The name EM implies that the technology uses time varying EM fields to probe the sediment properties. In general, low frequency EM technology has a source 'transmitter' that produces a time or frequency domain variation of a magnetic or electric field. The time varying field interacts with the electrical properties of the water and sea bottom sediments to produce induced electric currents. A receiver is used to detect and measure these induced currents. The sea bottom properties can then be estimated from the measurements by fitting the data to a simple model of the bottom. As in the electric sensor technology, geometry of the source and receiver is very important to get a good estimate of bottom conductivity properties.

An electrode resistivity array uses two source electrodes plus one or more sets of receiver electrodes. Resistivity techniques have been used in shallow exploration for more than 80 years however most recent applications are in the well logging industry. In the mid 80's, Valent, Mozley, and Corwin used an inverted Schlumberger array configuration as a towed array behind a ship to classify seafloor sediments[11].

Another type of ship towed system that has been used for sediment classification uses Controlled Source EM (CSEM) technology. One type developed by the Geological Survey of Canada more than 15 years ago consists of a towed magnetic source with following magnetic receivers. Woods Hole Oceanographic Institution uses an enhanced version of this system to support research on the continental shelf. A diagram of the basic deployment configuration is shown in figure 12. The transmitter is a horizontal magnetic source and the receivers are three horizontal magnetic sensors towed behind the transmitter at different offset distances. The system uses AC fields and produces data that can be related to seafloor electrical properties to roughly 20 meters in the sub-bottom.

Another CSEM technology that has gained significant use for the oil industry in the past ten years uses a towed electric source and an array of bottom mounted magnetic and electric receivers to investigate the electrical properties of the sub-bottom[12]. In this type survey system, a ship tows an electric source (two electrodes separated by some cabled distance) slightly above the bottom. Magnetic and electric field receiver systems are positioned on the bottom to record the data as the towed source is moved through the array of receivers. An illustration of the deployment technique is shown in figure 13. In this configuration the source to receiver offset varies significantly and supports a wide range of sub-bottom depth.

Another EM technique that is used in commercial applications and has been applied to marine surveys is the Airborne EM system. AEM technology has been used in prospecting for more than 40 years. In the 1980's the technology was applied to measure shallow marine bathymetry and bottom properties. This is also a CSEM type technique with a fixed transmitter and receiver pair configured as a vertical magnetic source and receiver pair (horizontal coils). The system is often towed under a helicopter as shown in figure 14. An AEM system can be used to measure water depth as well as bottom electrical properties for shallow water locations[13].

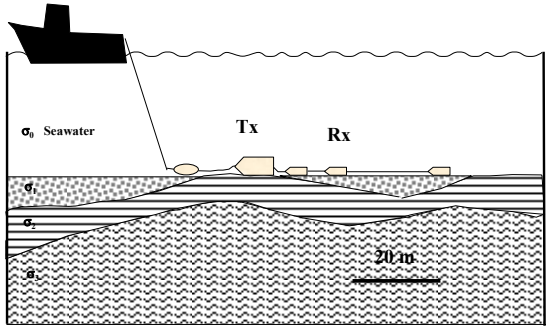
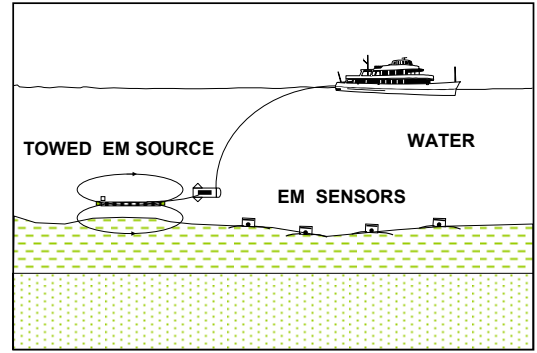

**Figure 12 Towed controlled source EM system for bottom characterization. (based on drawing from Rob Evans [2])**

**Figure 13 CSEM survey technique used in marine exploration of the sub-bottom.**

Early technology primarily centered on the use of electric field sensors. As technology and electronics have improved, both electric and EM sensor technologies have been incorporated as separate tools with specific applications. Trends in the electromagnetic measurement of seafloor sediment properties have been driven by a combination of technology innovation and commercial motivation. New applications in the oil industry have supported advances in both the survey equipment and survey technology to make large seabed surveys commercially viable. Future advances in the sensing technology will most likely be developed to support underwater applications on small autonomous systems.

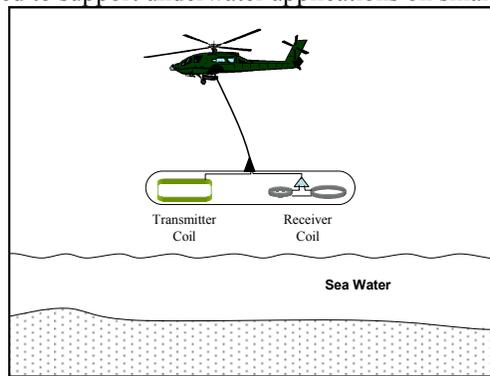

**Figure 14 AEM system concept for measuring water and bottom electrical properties.**

*Acoustic Techniques*

While the reflection and scattering of light is usually the best method for environmental sensing in air (humans rely primarily on our sense of sight), acoustic reflection and scattering provides a more effective means for remote sensing under water. Attenuation and water-column scattering prevents light from propagating more than a few 10's of feet. On the other hand, low frequency sound can travel 100's or even 1000's of miles in the ocean.(a reference here would be nice). Early unidirectional sonar techniques were developed to determine water depth (bathymetry) directly below a ship. Later advances enabled directional submarine detection and multidirectional bathymetry (multibeam echosounders). With large amounts of quality acoustic data available, several methods have been developed to classify seafloor sediments acoustically. Three broad categories for acoustic classification techniques are 1) Impedance methods, 2) Scattering Model methods, and 3) Acoustic Texture methods.

*Impedance methods*: Acoustic impedance $Z$ of a seafloor sediment is, by definition, $Z = \rho c$ where $\rho$ is the density and $c$ is the sound speed of the material. The reflection coefficient $R$ at the seafloor is,

$$R = \frac{Z - Z_0}{Z + Z_0},$$

where $Z_0$ is the acoustic impedance of the seawater. So the reflection coefficient increases with the sediment impedance, and therefore, the impedance can be measured by the strength (loudness) of sonar echoes. Sediment properties that may be inferred by the impedance measurements include density, porosity, and shear strength. This method may be used with fathometers, subbottom profilers, and chirp sonars. Two product examples are RoxAnn (derives reflection coefficient from ratio of first and second bottom echoes) and NRL's Acoustic Sediment Classification System[1] (ASCS), which can use both fathometer and chirp sonar systems.

*Scattering Model methods*: Acoustic seafloor scattering models (for example see [14,15]) predict the intensity of scattered sound as a function of incident and scattering angles. Scattering back in the incidence direction is known as backscatter. With multibeam echosounders measurements of backscattered sound intensity levels can be made at many backscattering (or grazing) angles. By fitting the model to the multibeam data, optimal model parameters are found for impedance, acoustic roughness (on a scale of 1-meter), and bulk sediment (volume) interaction. NRL has developed SediMap®[16,17], a software package to characterize the seafloor sediment using multibeam sonar. Mean-grain-size estimates based on SediMap impedance calculations correlate with ground truth with a correlation coefficient of about 0.6-0.7 (about the same as the correlation between mean-grain-size and impedance measurements from core samples).

*Acoustic Texture methods*: Higher frequency sonars can produce sufficient resolution as to be able to determine acoustic "graininess", or time fluctuations in the echo intensities. This can be visually demonstrated with images from sidescan sonar. Quester Tangent[18] and, more recently, Qinetic have developed products to analyze statistical properties in high-resolution imagery to separate regions containing different sediments. NRL's Data Fusion from Acoustic Backscattering from the Seafloor (DFABS) uses an analog to a magnetic system (Potts model) for rapid image segmentation of sidescan images. These methods have proven successful in recognizing areas with different sediment types, however, they are less relilable in classifying the sediment (in part due to the lack of penetration at high frequencies). Additional information is needed for reliable characterization.

Recent R&D trends aim to combine two or more of the above methods to obtain the best characterization of the seafloor (data fusion). DFABS is under development and will combine sidescan image segmentation with ASCS for identifying sediment regions and their properties. Further trends involve making use of fleet sensors[19,20] (Through-The-Sensor technologies) and synthetic aperture sonar (SAS) and increasing capabilities for near real-time processing at sea[9].

## V. SUMMARY AND FUTURE DIRECTIONS

In the recent history of sediment sampling techniques, there has been a pronounced addition of remote techniques to complement direct measurement of properties. 40 years ago engineers and scientists measured surficial sediment properties in shallow areas with cores and grab samplers. Due to the limited coverage and expense of these point measurements, techniques have evolved that provide areal coverage using remote acoustic and electromagnetic sensors. These new sensing techniques measure few properties directly; however, additional properties are derived from empirical data. Calibration of the empirical data is performed using core and grab sampling in selected areas. Data fusion techniques are being used increasingly to merge data collected by different techniques to further expand areal coverage. Future sensing techniques will begin to address 3D characterization of sediments in shallow coastal areas.

## VI. ACKNOWLEDGMENTS